\documentstyle[aps,prl,epsf]{revtex}

\bibstyle{unsrt}
\begin{document}
\draft

% \preprint{WUHEP-98-18}

\title{Complex periodic potentials with real band spectra}

\author{Carl M. Bender$^1$, Gerald V. Dunne$^2$, and Peter N. Meisinger$^1$}
\address{${}^1$Department of Physics, Washington University, St. Louis MO
63130, USA}
\address{${}^2$Department of Physics, University of Connecticut, Storrs CT
06269, USA}

\date{\today}
\maketitle

\begin{abstract}
This paper demonstrates that complex ${\cal PT}$-symmetric periodic potentials
possess real band spectra. However, there are significant qualitative
differences in the band structure for these potentials when compared with
conventional real periodic potentials. For example, while the potentials $V(x)=i
\sin^{2N+1}(x)$ $(N=0,~1,~2,~\ldots)$ have infinitely many gaps, at the band
edges there are periodic wave functions but no antiperiodic wave functions.
Numerical analysis and higher-order WKB techniques are used to establish these
results.
\end{abstract}
\vskip .5cm

For a quantum mechanical model having a periodic potential the Schr\"odinger
equation is
\begin{eqnarray}
-\psi''(x)+V(x)\psi(x)=E\psi(x),
\label{e1}
\end{eqnarray}
where the potential $V(x)$ is periodic with period $P$:
\begin{eqnarray}
V(x+P)=V(x).
\label{e2}
\end{eqnarray}
In conventional treatments of Eq.~(\ref{e1}) \cite{KITTEL,HILL} the periodic
potential $V(x)$ is assumed to be real. Imposing the condition that the wave
function $\psi(x)$ be bounded leads to a real spectrum consisting of continuous
bands separated by gaps. There is an infinite number of bands and gaps, except
for the special family of so-called {\it finite-gap} potentials such as the
Lam\'e potentials \cite{LAME}.

In this note we extend the conventional analysis to include the case of complex
periodic potentials. We find that complex periodic potentials having ${\cal PT}$
symmetry exhibit {\it real} band spectra, despite the non-Hermitian character of
the Schr\"odinger equation (\ref{e1}). (Here, ${\cal P}$ represents parity
reflection and ${\cal T}$ represents time reversal.) Potentials having this
symmetry satisfy
\begin{eqnarray}
[V(-x)]^*=V(x).
\label{e3}
\end{eqnarray}
Examples of such potentials are $i\sin x$, $i\sin^3(x)$, and $e^{ix}$. In
addition to the property that these potentials have real spectra, the band
structure displays several novel features that are strikingly different from the
case of real periodic potentials.

The work reported here was motivated by recent investigations of non-Hermitian
${\cal PT}$-symmetric Hamiltonian models having real discrete spectra. One such
class of models is defined by the Hamiltonian \cite{PRL}
\begin{eqnarray}
H=p^2+x^2(ix)^\epsilon\quad(\epsilon\geq0).
\label{e4}
\end{eqnarray}
Despite the lack of conventional Hermiticity, the spectrum of this Hamiltonian
is real, positive, and discrete; each of the energy levels increases as a
function of increasing $\epsilon$. It has been observed that the reality of the
spectrum is a consequence of ${\cal PT}$ symmetry, which is a weaker condition
than Hermiticity. This observation has also been used to construct new classes
of quasi-exactly solvable quantum theories \cite{QES} and to study new kinds of
symmetry breaking in quantum field theory \cite{PARITY,SUP}. There have been
many other instances of non-Hermitian ${\cal PT}$-invariant Hamiltonians in
physics. Energies of solitons on a {\it complex} Toda lattice have been found to
be real \cite{HOLLOW}. Hamiltonians rendered non-Hermitian by an imaginary
external field have been used to study population biology \cite{Nelson+Shnerb}
and to study delocalization transitions such as vortex flux-line depinning in
type-II superconductors \cite{Hatano+Nelson}. In these last two cases, initially
real eigenvalues bifurcate into the complex plane due to the increasing external
field, indicating the growth of populations or the unbinding of vortices.

We begin by summarizing the standard Floquet analysis of the Schr\"odinger
equation (\ref{e1}) for the case where $V(x)$ is real and periodic \cite{HILL}.
We define a {\it fundamental pair} of linearly independent solutions $u_1(x)$
and $u_2(x)$ satisfying the initial conditions
\begin{eqnarray}
u_1(0)&=&1,\quad u_1'(0)=0;\nonumber\\
u_2(0)&=&0,\quad u_2'(0)=1.
\label{e5}
\end{eqnarray}
Any solution $\psi(x)$ to Eq.~(\ref{e1}) is a linear combination of $u_1(x)$ and
$u_2(x)$. It is then a straightforward algebraic exercise to show that $\psi(x)$
is bounded provided that the {\it discriminant} $\Delta(E)$, which is defined by
\begin{eqnarray}
\Delta(E)\equiv u_1(P)+u_2'(P),
\label{e6}
\end{eqnarray}
satisfies the constraint
\begin{eqnarray}
-2\leq\Delta(E)\leq2.
\label{e7}
\end{eqnarray}

\begin{figure*}[p]
\vspace{3.2in}
\includegraphics{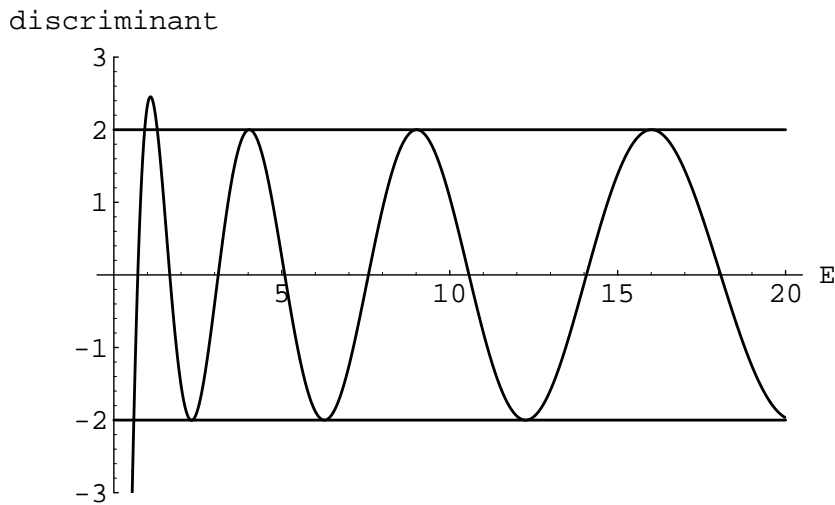}
\caption{The discriminant $\Delta(E)$ plotted as a function of $E$ for the real
periodic potential $V(x)=\sin(x)$. Although it cannot be seen in the figure,
all local maxima lie above the line $\Delta=2$ and all local minima lie below
the line $\Delta=-2$. The regions of energy $E$ for which $|\Delta|\leq2$
correspond to bands and the regions where $|\Delta|>2$ correspond to gaps. There
are infinitely many gaps and these gaps become exponentially narrow as $E$
increases.}
\label{f1}
\end{figure*}

To illustrate the features of the discriminant we consider a typical periodic
potential, $V(x)=\sin(x)$, for which the period $P=2\pi$. In Fig.~\ref{f1} we
plot $\Delta(E)$ as a function of $E$. Note that $\Delta(E)$ is oscillatory and
is well approximated by the function $2\cos(\pi\sqrt{E})$ for large $E$. The
crucial feature of $\Delta(E)$, which cannot be seen from this plot, is that its
graph crosses the lines $\pm2$ an infinite number of times; each of the maxima
of $\Delta(E)$ lies above $2$ and each of the minima lies below $-2$. The
regions of energy for which $|\Delta(E)|\leq2$ are called {\it bands} and the
regions of energy for which $|\Delta(E)|>2$ are called {\it gaps}. The gap size
decreases exponentially as a function of $E$. The band edges at which $\Delta(E)
=2$ correspond to periodic solutions to Eq.~(\ref{e1}), $\psi(x)=\psi(x+P)$, and
the band edges at which $\Delta(E)=-2$ correspond to antiperiodic solutions
$\psi(x)=-\psi(x+P)$.

Now consider the calculation of the discriminant for the case of a {\it complex}
periodic potential $V(x)$. In general, a complex periodic potential will have no
bounded solutions because the discriminant is typically complex. However, for
complex ${\cal PT}$-symmetric periodic potentials, one can easily show that the
discriminant $\Delta(E)$ is real when $E$ is real. The ${\cal PT}$ symmetry is
crucial here; for a potential that is not ${\cal PT}$ symmetric [one that does
not satisfy Eq.~(\ref{e3})], the discriminant is complex for all values of $E$.
Having established that complex ${\cal PT}$-symmetric periodic potentials have
real discriminants, we can then apply the criterion in Eq.~(\ref{e7}) to locate
the real energy bands within which the corresponding wave function $\psi(x)$ is
a bounded function.

\begin{figure*}[p]
\vspace{3.2in}
\includegraphics{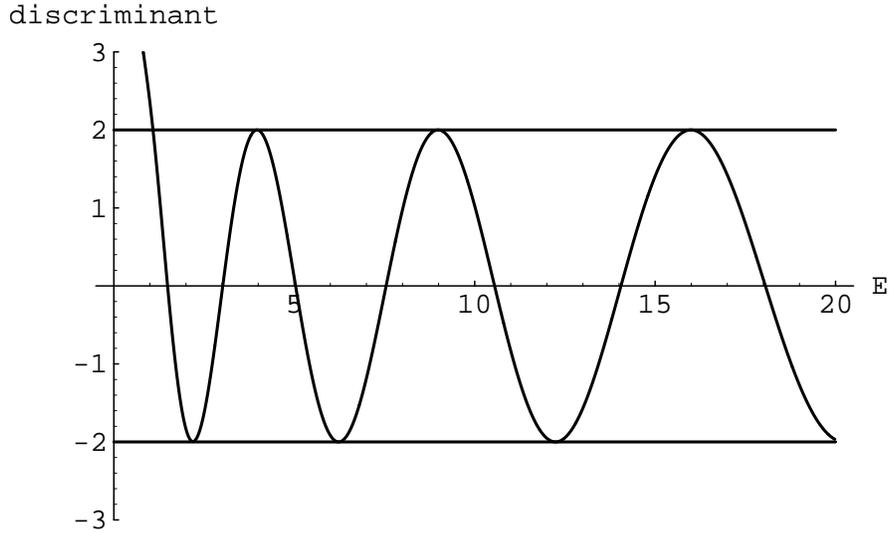}
\caption{The discriminant $\Delta(E)$ plotted as a function of $E$ for the
complex periodic potential $V(x)=i\sin(x)$. On the scale of this figure for
$E>2$ it is not possible to see any difference between this figure and Fig.~1.
However, although it cannot be seen in this figure, all local maxima lie above
the line $\Delta=2$ and all local minima lie above the line $\Delta=-2$. This
behavior is distinctly different from the generic behavior in Fig.~1 exhibited
by the real periodic potential $V(x)=\sin x$.}
\label{f2}
\end{figure*}

\begin{figure*}[p]
\vspace{3.2in}
\includegraphics{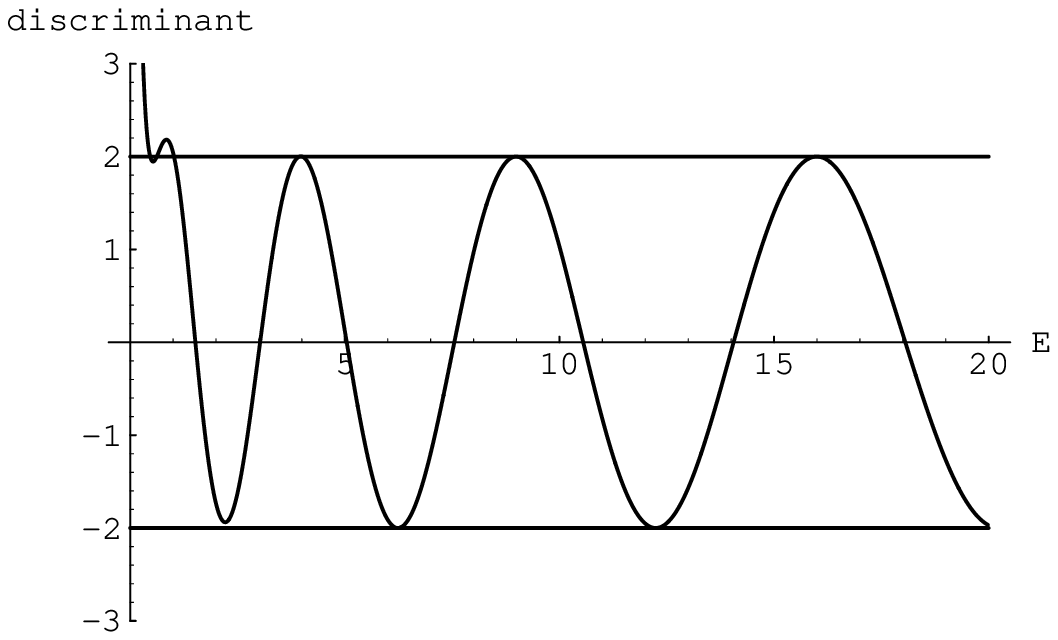}
\caption{The discriminant $\Delta(E)$ for the complex periodic potential
$V(x)=i\sin^3(x)$. All local maxima lie above the line $\Delta=2$ and all local
minima lie above the line $\Delta=-2$. Note the appearance of a new local
minimum near $E=0.5$.}
\label{f3}
\end{figure*}

\begin{figure*}[p]
\vspace{3.2in}
\includegraphics{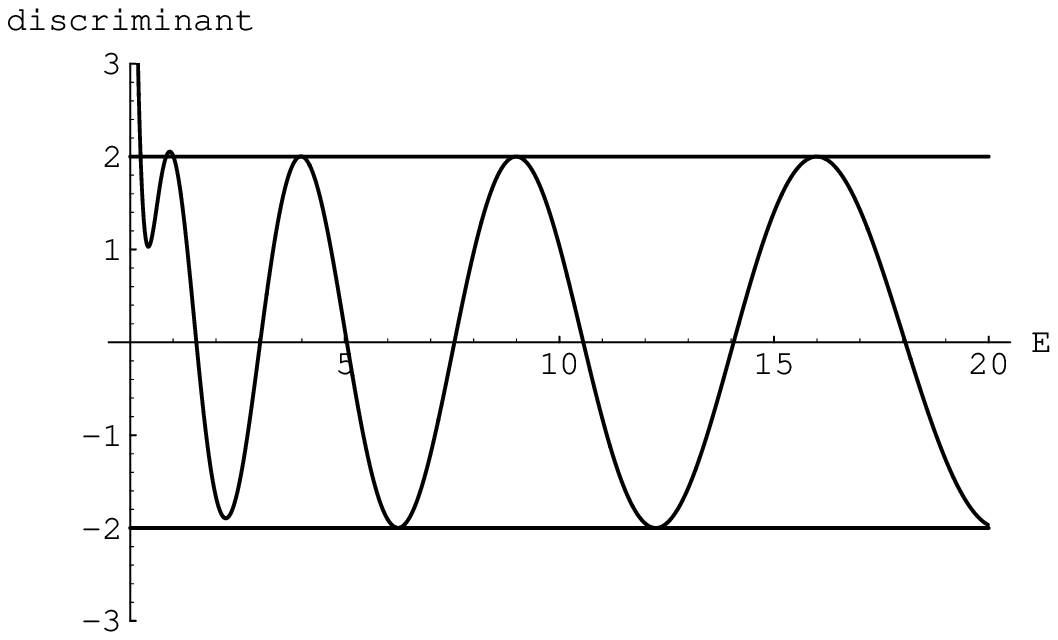}
\caption{The discriminant $\Delta(E)$ for the complex periodic potential
$V(x)=i\sin^5(x)$. All local maxima lie above the line $\Delta=2$ and all local
minima lie above the line $\Delta=-2$.}
\label{f4}
\end{figure*}

\begin{figure*}[p]
\vspace{3.2in}
\includegraphics{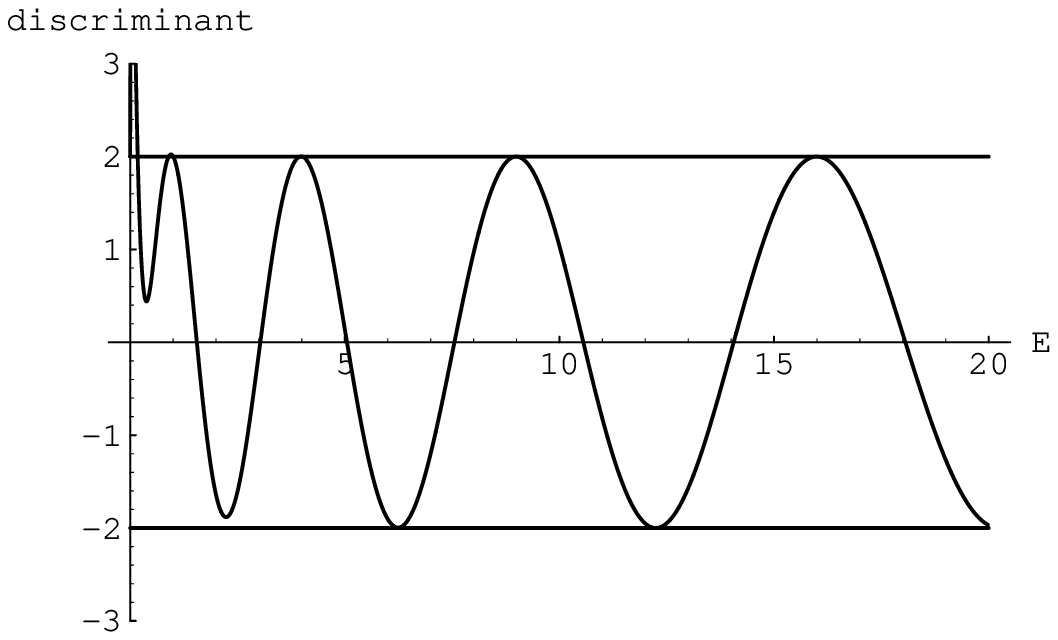}
\caption{The discriminant $\Delta(E)$ for the complex periodic potential
$V(x)=i\sin^7(x)$. All local maxima lie above the line $\Delta=2$ and all local
minima lie above the line $\Delta=-2$.}
\label{f5}
\end{figure*}

\begin{figure*}[p]
\vspace{3.2in}
\includegraphics{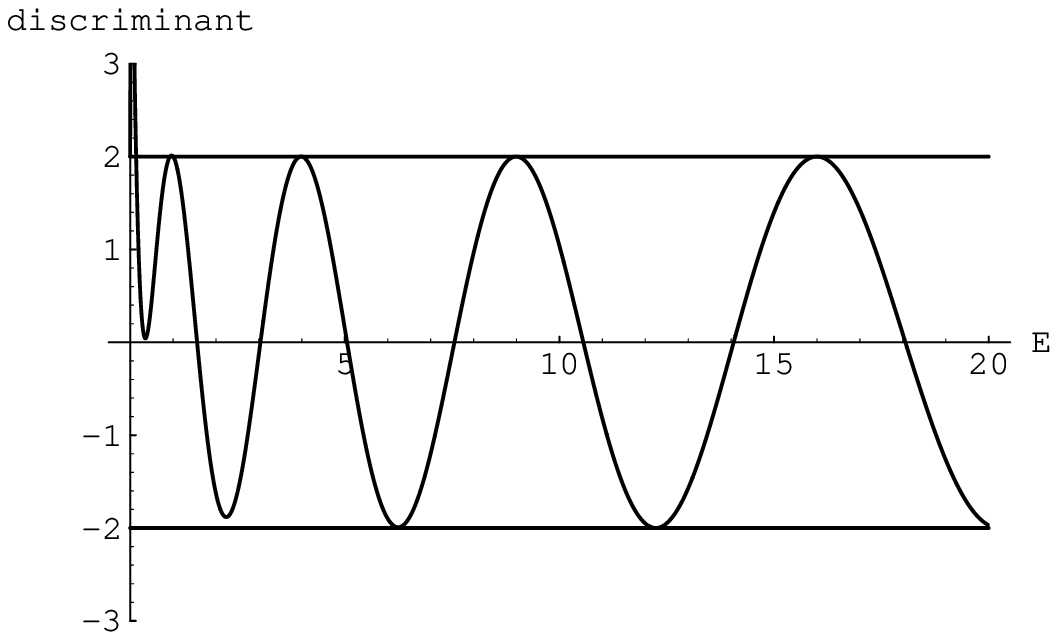}
\caption{The discriminant $\Delta(E)$ for the complex periodic potential
$V(x)=i\sin^9(x)$. All local maxima lie above the line $\Delta=2$ and all local
minima lie above the line $\Delta=-2$.}
\label{f6}
\end{figure*}

\begin{figure*}[p]
\vspace{3.2in}
\includegraphics{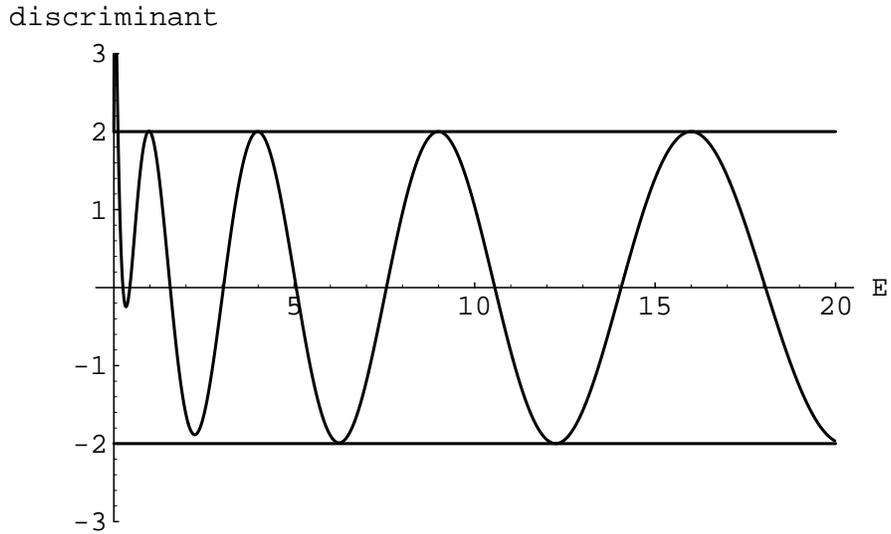}
\caption{The discriminant $\Delta(E)$ for the complex periodic potential
$V(x)=i\sin^{11}(x)$. All local maxima lie above the line $\Delta=2$ and all
local minima lie above the line $\Delta=-2$.}
\label{f7}
\end{figure*}

We have computed the discriminants for the class of complex ${\cal
PT}$-symmetric periodic potentials
\begin{eqnarray}
V(x)=i\sin^{2N+1}(x)\quad(N=0,~1,~2,~\ldots).
\label{e8}
\end{eqnarray}
In Figs.~\ref{f2}-\ref{f7} we plot the discriminants for the cases $N=0,~1,~
\ldots,~5$. While these plots superficially resemble Fig.~\ref{f1} for large
$E$, they exhibit new and intriguing features that are significantly different
from the case of a real periodic potential. The most obvious new feature is the
appearance for $N\geq1$ of a local minimum of $\Delta(E)$ between $-2$ and $2$.
Such a dip is rigorously forbidden in the case of real periodic potentials
\cite{HILL}.

The most dramatic differences between the discriminants for the complex ${\cal
PT}$-symmetric periodic potentials in Eq.~(\ref{e8}) and real periodic
potentials cannot be easily seen in the figures. We have performed a careful
numerical study of the local minima and maxima of $\Delta(E)$. Our study reveals
that {\it none of the local minima lies below} $-2$. This shows that there are
{\it no antiperiodic solutions} $\psi(x)$ to the Schr\"odinger equation
(\ref{e1}). Nevertheless, all of the local maxima of $\Delta(E)$ lie above $2$.
Hence, there are an infinite number of band gaps in the spectrum and the
band-edge wave functions are periodic.

\vfill\eject
$$\phantom{XXXXXXXXXXXXXXXXXXXXXXXXXXXXXXXXX}$$

\vfill\eject
$$\phantom{XXXXXXXXXXXXXXXXXXXXXXXXXXXXXXXXX}$$

To perform this numerical analysis it is necessary to locate the positions of
the local minima and maxima of $\Delta(E)$ to extremely high accuracy. This can
be done using WKB methods \cite{BO}. We take the energy $E$ to be large $(E>>1)$
and define a small parameter $\epsilon$ by
\begin{eqnarray}
\epsilon={1\over\sqrt{E}}.
\label{e9}
\end{eqnarray}
Then we make an exponential {\it ansatz} for the wave function $\psi(x)$:
\begin{eqnarray}
\psi(x)=\exp\left[{i\over\epsilon}\sum_{n=0}^{\infty}\epsilon^nQ_n(x)\right].
\label{e10}
\end{eqnarray}

Substituting $\psi(x)$ in Eq.~(\ref{e10}) into the Schr\"odinger equation
(\ref{e1}) gives a recursion relation for the functions $Q_n(x)$:
\begin{eqnarray}
1-[Q_0'(x)]^2&=&0,\nonumber\\
iQ_0''(x)-2Q_0'(x)Q_1'(x)&=&0,\nonumber\\
iQ_1''(x)-2Q_0'(x)Q_2'(x)-[Q_1'(x)]^2-V(x)&=&0,\nonumber\\
iQ_{n-1}''(x)-\sum_{j=0}^nQ_j'(x)Q_{n-j}'(x)&=&0\quad(n\geq3).
\label{e11}
\end{eqnarray}
The solution to these equations is
\begin{eqnarray}
Q_0(x)&=&\pm x,\nonumber\\
Q_1(x)&=&0,\nonumber\\
Q_2(x)&=&\mp{1\over2}\int_0^xdt\,V(t),\nonumber\\
Q_3(x)&=&-{i\over4}\left[V(x)-V(0)\right],\nonumber\\
Q_4(x)&=&\pm{1\over8}\left(V'(x)-V'(0)-\int_0^xdt\,V^2(t)\right),\nonumber\\
Q_5(x)&=&{i\over16}\left[V''(x)-V''(0)-2V^2(x)+2V^2(0)\right],\nonumber\\
Q_6(x)&=&\mp{1\over32}\left(V'''(x)-V'''(0)-5V(x)V'(x)+5V(0)V'(0)
+\int_0^xdt\,[2V^3(t)-V(t)V''(t)]\right),
\label{e12}
\end{eqnarray}
and so on. Note that $Q_n(x)$ is normalized so that $Q_n(0)=0$. In general, the
formula for $Q_n'(x)$ is
\begin{eqnarray}
Q_n'(x)=\pm{1\over2}\left[iQ_{n-1}''(x)-\sum_{j=1}^{n-1}Q_j'(x)Q_{n-j}'(x)
\right]\quad(n\geq3).
\label{e13}
\end{eqnarray}

In order to obtain a WKB formula for the discriminant $\Delta(E)$ in
Eq.~(\ref{e6}) we need to evaluate $Q_n(x)$ at $x=P$. The periodicity of the
potential $V(x)$ simplifies the results considerably; when $n$ is odd,
$Q_n(P)=0$ and when $n$ is even, only the integrals in Eq.~(\ref{e12}) remain:
\begin{eqnarray}
Q_0(P)&=&\pm P,\nonumber\\
Q_2(P)&=&\mp{1\over2}\int_0^Pdt\,V(t),\nonumber\\
Q_4(P)&=&\mp{1\over8}\int_0^Pdt\,V^2(t),\nonumber\\
Q_6(P)&=&\mp{1\over32}\int_0^Pdt\,[2V^3(t)-V(t)V''(t)],
\label{e14}
\end{eqnarray}
and so on.

The WKB formula for the discriminant is particularly simple when the potential
$V(x)$ is ${\cal PT}$ symmetric:
\begin{eqnarray}
\Delta(E)=2\cos\left[{1\over\epsilon}\sum_{n=0}^{\infty}\epsilon^{2n}Q_{2n}(P)
\right].
\label{e15}
\end{eqnarray}
One obtains the same formula for potentials that are real and symmetric under
parity ${\cal P}$.

For the complex ${\cal PT}$-symmetric potentials $V(x)$ in (\ref{e8}) the WKB
formula for the discriminant in Eq.~(\ref{e15}) is
\begin{eqnarray}
\Delta(E)=2\cos\left[{2\pi\over\epsilon}+{\Gamma(2N+3/2)\sqrt{\pi}\over
4(2N+1)!}\,\epsilon^3+{(2N+1)\Gamma(2N+1/2)\sqrt{\pi}\over32(2N)!}\,\epsilon^5
+\cdots\right].
\label{e16}
\end{eqnarray}
A similar WKB formula exists for the real odd-parity potentials $V(x)=
\sin^{2N+1}(x)$:
\begin{eqnarray}
\Delta(E)=2\cos\left[{2\pi\over\epsilon}-{\Gamma(2N+3/2)\sqrt{\pi}\over
4(2N+1)!}\,\epsilon^3-{(2N+1)\Gamma(2N+1/2)\sqrt{\pi}\over32(2N)!}\,\epsilon^5
+\cdots\right].
\label{e17}
\end{eqnarray}

We can illustrate the extreme accuracy of these WKB approximations by comparing
them with numerical computations of the discriminant. For example, for $V(x)=i
\sin^3(x)$ at $\epsilon=0.2$ (which corresponds to $E=25$) we find numerically
that $\Delta(25)=1.9999960002$. The first three orders of the WKB approximation
taken from Eq.~(\ref{e16}) give $\Delta(25)=2.0$, $\Delta(25)=1.9999961447$, and
$\Delta(25)=1.9999960046$. Similarly, for $V(x)=\sin^3(x)$ at $\epsilon=0.2$ we
find numerically that $\Delta(25)= 1.9999959937$. The first three orders of the
WKB approximation taken from Eq.~(\ref{e17}) give the same values: $\Delta(25)=
2.0$, $\Delta(25)=1.9999961447$, $\Delta(25)=1.9999960046$.

Despite this impressive precision, the WKB formulas (\ref{e16}) and (\ref{e17})
cannot be used directly to answer the crucial question of whether there are band
gaps because these approximations to the discriminant $\Delta(E)$ never cross
the values $\pm2$. The reason for this inadequacy of the WKB approximation is
that the differences $|\Delta_{\rm max}-2|$ and $|\Delta_{\rm min}+2|$ are
exponentially small when $E>>1$. Therefore, these differences are subdominant
with respect to the WKB asymptotic series and are not accessible to any order in
powers of $\epsilon$. Indeed, the WKB series can only provide information about
quantities with an {\it algebraically} small error, and not an exponentially
small error. We emphasize that the WKB approximation has this shortcoming only
at the maxima and minima of the approximation. At other points any exponential
discrepancy is completely negligible compared with algebraic errors.

The WKB series is still an extremely useful ingredient in the numerical search
for zeros of $\Delta(E)\pm2$. (These zeros are the dividing points between bands
and gaps.) Our procedure is first to find the energies at which there are maxima
and minima of the WKB approximation to the discriminant and then to evaluate,
with high numerical precision, the actual value of the discriminant in a tiny
neighborhood of each of these points. By doing this we can determine whether or
not the discriminant $\Delta(E)$ crosses the lines $\pm2$.

For the real periodic potentials $V(x)=\sin^{2N+1}(x)$ our procedure confirms
the rigorous theoretical result that every maximum of the discriminant lies
above $2$ and every minimum lies below $-2$. Consider, for example, the
potential $V(x)=\sin(x)$. From Fig.~1, it is clear that the first maximum lies
above $2$. The first minimum occurs at $E=2.3138$, where the discriminant has
the value $-2.0038787$. The second maximum occurs at $E=4.0336$, where the
discriminant is $2.000007$. Similar behavior is found for the other potentials
in the class $V(x)=\sin^{2N+1}(x)$. In stark contrast, for the potentials
$V(x)=i\sin^{2N+1}(x)$, while the maxima of the discriminant lie above $+2$, the
minima of the discriminant lie above $-2$. Thus, for these potentials there are
no antiperiodic wave functions. As an example, lengthy and delicate numerical
analysis verifies that for the potential $V(x)=i\sin(x)$ the first three maxima
of the discriminant $\Delta(E)$ are located at $E=3.9664284$, $E=8.9857320$, and
$E=15.99206621346$. The value of the discriminant $\Delta(E)$ at these energies
is $2.000007$, $2.00000000000069$, and $2.000000000000000000015$. The first two
minima of the discriminant are located at $E=2.1916$ and $E=6.229223$ and at
these energies $\Delta(E)$ has the values $-1.9953386$ and $-1.99999999527$.
Similar behavior is found for the other potentials in the class (\ref{e8}).

We conclude by pointing out that from the expressions for $Q_n(P)$ in
Eq.~(\ref{e14}) the WKB series truncates if the potential is a polynomial in
$e^{ix}$. For example, for the complex ${\cal PT}$-symmetric periodic potential
$V(x)=e^{ix}$ the WKB series in Eq.~(\ref{e15}) truncates after the first term
because $Q_n(2\pi)$ vanishes for $n\geq1$. For this case the WKB approximation
is exact and the discriminant is given by
\begin{eqnarray}
\Delta(E)=2\cos(2\pi\sqrt{E}).
\label{e18}
\end{eqnarray}
One can verify this result directly by solving the Schr\"odinger equation
(\ref{e1}) for this potential exactly; the solution is a Bessel function:
$\psi(x)=J_{2\sqrt{E}}(2e^{ix/2})$.

\section* {ACKNOWLEDGEMENT}
\label{s6}

We are grateful to the U.S.~Department of Energy for financial support.

\end{document}